\documentclass{mem}
\usepackage{natbib}\usepackage{txfonts}\usepackage{balance}
\usepackage{graphicx}
\usepackage[a4paper,breaklinks,dvipdfm]{hyperref}
\idline{?}{1}
\begin{document}
\def\teff{$T\rm_{eff }$}
\def\kms{$\mathrm {km s}^{-1}$}

\title{
Stellar populations\\ in tidally stirred dwarf galaxies
}

   \subtitle{}

\author{
E. L. {\L}okas\inst{1}, K. Kowalczyk\inst{2} \and S. Kazantzidis\inst{3}
          }

  \offprints{E. L. {\L}okas}

\institute{
Nicolaus Copernicus Astronomical Center, Bartycka 18, 00-716 Warsaw, Poland, \email{lokas@camk.edu.pl}
\and
Warsaw University Observatory, Al. Ujazdowskie 4, 00-478 Warsaw, Poland
\and
Center for Cosmology and Astro-Particle Physics and Department of Physics and Department of Astronomy,
    The Ohio State University, Columbus, OH 43210, USA
}

\authorrunning{{\L}okas et al.}

\titlerunning{Stellar populations in tidally stirred dwarf galaxies}

\abstract{
Using $N$-body simulations we study the evolution of separate stellar populations in dwarf galaxies
in the context of the tidal stirring scenario for the formation of dwarf spheroidal (dSph) galaxies
in the Local Group. The dwarf galaxies,
initially composed of a stellar disk and a dark matter halo, are placed on seven different orbits
around the Milky Way. The stars are divided into two populations, within and outside the half-light
radius, and their positions are followed for 10 Gyr. We find that the populations retain different
density distributions even over such long timescales. Some of the stars of the outer population migrate to the
central part of the dwarf forming an extended core while the stars of the inner population develop a tail
in the outer parts. In addition, the outer population is more heavily stripped by tidal forces from the
Milky Way and may become subdominant at all radii on tight enough orbits. We conclude that
the tidal stirring model is fully compatible with the presence of multiple stellar populations in
dSph galaxies.
}
\maketitle{}

\section{Introduction}

Dwarf spheroidal (dSph) galaxies of the Local Group provide unique laboratory for studying the processes of
galaxy formation and evolution. Their relative proximity has recently enabled detailed probing of density
distributions and kinematics, as well as metallicities of their stellar components. These observational studies
have revealed the presence of multiple stellar populations in the dwarfs, differing by metallicity and age.
The populations thus identified were shown to be distributed differently in space and possess various velocity
distributions as quantified e.g. by their velocity dispersion profiles (Tolstoy et al. 2004; Koch et al. 2006;
Battaglia et al. 2008; Kirby et al. 2011; de Boer et al. 2012).
The presence of multiple stellar populations has been invoked as a possible explanation for anomalous
velocity dispersion profiles observed in some dSph galaxies (Ibata et al. 2006; McConnachie et al. 2007).
It has also recently proved useful in constraining the density slope of dSph galaxies (Walker \& Pe\~narrubia 2011).

\begin{table*}
\caption{Parameters of the simulated dwarfs}
\label{parameters}
\begin{center}
\begin{tabular}{ccccccccc}
\hline
\\
Orbit & $r_{\rm apo}$  & $r_{\rm peri}$  & $r_{\rm apo}/r_{\rm peri}$ & $T_{\rm orb}$  &
 $n_{\rm peri}$ & $M_V$ & $r_{1/2}$ & $\mu_V$ \\
     &  [kpc] &  [kpc] &   & [Gyr] &   &  [mag] &  [kpc]  & [mag arcsec$^{-2}$] \\
\hline
\\
O1  &    125  &     25    &   \ \   5    &  2.09  &  5 &$   -11.5$ & 0.32 & 23.4 \\
O2  & \ \ 85  &     17    &   \ \   5    &  1.28  &  8 &$   -10.0$ & 0.35 & 24.8 \\
O3  &    250  &     50    &   \ \   5    &  5.40  &  2 &$   -12.4$ & 0.62 & 23.8 \\
O4  &    125  & \ \ 12.5  &   $\,  10$   &  1.81  &  6 &$\ \ -9.7$ & 0.25 & 24.7 \\
O5  &    125  &     50    & \ \ \ \ 2.5  &  2.50  &  4 &$   -12.3$ & 0.41 & 22.9 \\
O6  & \ \ 80  &     50    & \ \ \ \ 1.6  &  1.70  &  6 &$   -12.2$ & 0.43 & 23.3 \\
O7  &    250  & \ \ 12.5  &   $\,  20$   &  4.55  &  2 &$   -11.7$ & 0.39 & 23.5 \\
\hline
\end{tabular}
\end{center}
\end{table*}

Substantial progress has also been achieved in theoretical modelling of the formation of dSph galaxies. One
promising scenario for their origin invokes tidal evolution in the vicinity of more massive galaxies like the
Milky Way or Andromeda (Mayer et al. 2001). The tidal stirring model envisions the present-day dSph galaxies as
products of the repeated action of the tidal force from the host galaxy on a late-type progenitor accreted
a few Gyr ago, thus explaining naturally the observed morphology-density relation in the Local Group. The process
can be reliably modelled using $N$-body simulations which demonstrate that such evolution is indeed feasible over
reasonable timescales (Klimentowski et al. 2009; Kazantzidis et al. 2011). The transformation from disks to spheroids
is accompanied by randomization of stellar orbits and strong mass loss due to tidal stripping.

The morphological evolution typically involves a bar-like stage ({\L}okas et al. 2012)
and therefore a question arises what effect the
instability associated with bar formation may have on the properties of stellar populations present in the
dwarf galaxy. Since most of the stars in dSph galaxies are old, they have probably formed before the dwarf's
progenitor was accreted by the host. Although some star formation may occur as a result of tidal stirring,
the amount of new stars is probably not large as the gas reservoir is likely to be depleted quickly by ram pressure
stripping (Mayer et al. 2007).
In this work we address the question whether the separate stellar populations present in the
progenitor prior to accretion are likely to survive until the present time. Such considerations may put
some constraints on the amount of tidal stirring experienced by dSph galaxies in our vicinity.

\begin{figure*}[t!]
\resizebox{\hsize}{!}{\includegraphics[clip=true]{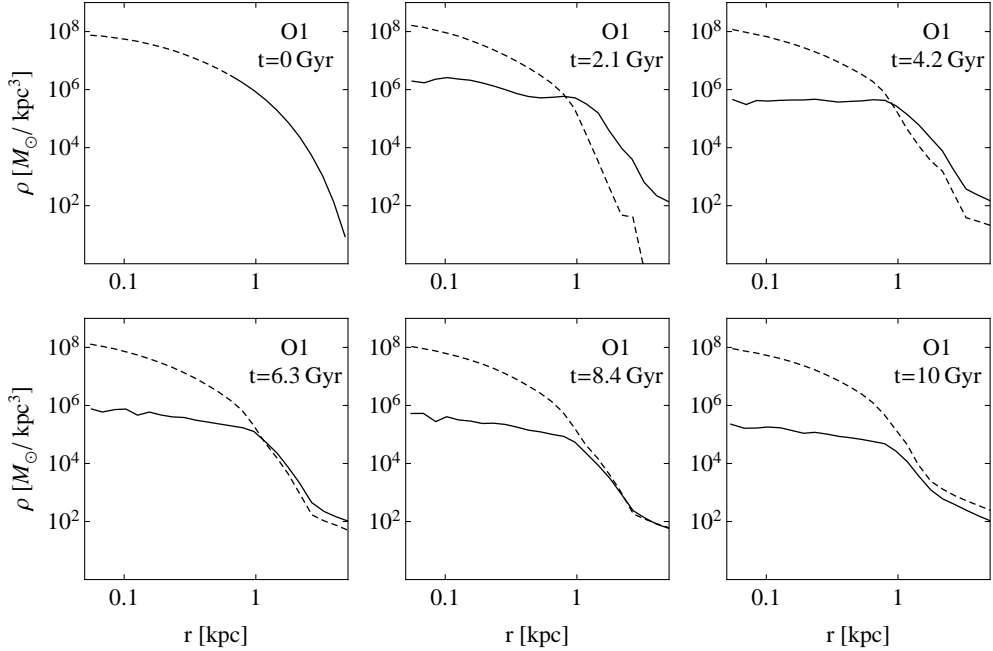}}
\caption{\footnotesize
Evolution of the density profiles of stars in the inner Population 1 (dashed lines) and the outer Population 2
(solid lines) for the dwarf galaxy on orbit O1.
The populations are initially divided by the half-light radius (upper left panel). The panels
(from the upper left to the lower right) show the density profiles at subsequent apocenters of the orbit
and at the final stage. The times of the measurements are given in each panel.
}
\label{orbit1}
\end{figure*}

\section{The simulations}

For the present study we used a subset of tidal stirring simulations described in detail in Kazantzidis et al. (2011)
and {\L}okas et al. (2011).
Our dwarf galaxy was initially composed of an exponential stellar disk of mass $2 \times 10^7 M_\odot$ embedded in
a dark matter halo of virial mass $10^9 M_\odot$ and an NFW (Navarro et al. 1996) density profile with concentration
$c=20$. The $N$-body realization of the dwarf contained $1.2 \times 10^6$ stars and $10^6$ dark matter particles.
The dwarf galaxy was placed on seven different orbits (initially at the apocenter)
around a host galaxy with the present-day properties of the
Milky Way (Widrow \& Dubinski 2005) and its evolution was followed for 10 Gyr with the PKDGRAV code (Stadel 2001).

The orbital parameters of the simulations are listed in Table~\ref{parameters}. Columns 2-6 of the Table
give respectively
the apocenter $r_{\rm apo}$, the pericenter $r_{\rm peri}$, the ratio of the two distances, the orbital time
and the number of pericenters passed during the 10 Gyr of evolution. The orbits cover a large range of sizes
and eccentricities resulting in different amounts of tidal force experienced by the dwarf galaxy. The last three
columns of Table~\ref{parameters} list the observational parameters of the dwarfs at the final stage as
they would be measured by an imaginary observer at the center of the Milky Way: the absolute magnitude, the
projected half-light radius and the central surface brightness ({\L}okas et al. 2011). The parameters fall in the
ranges occupied by the classical dSph galaxies in the vicinity of the Milky Way.

\begin{figure*}[t!]
\resizebox{\hsize}{!}{\includegraphics[clip=true]{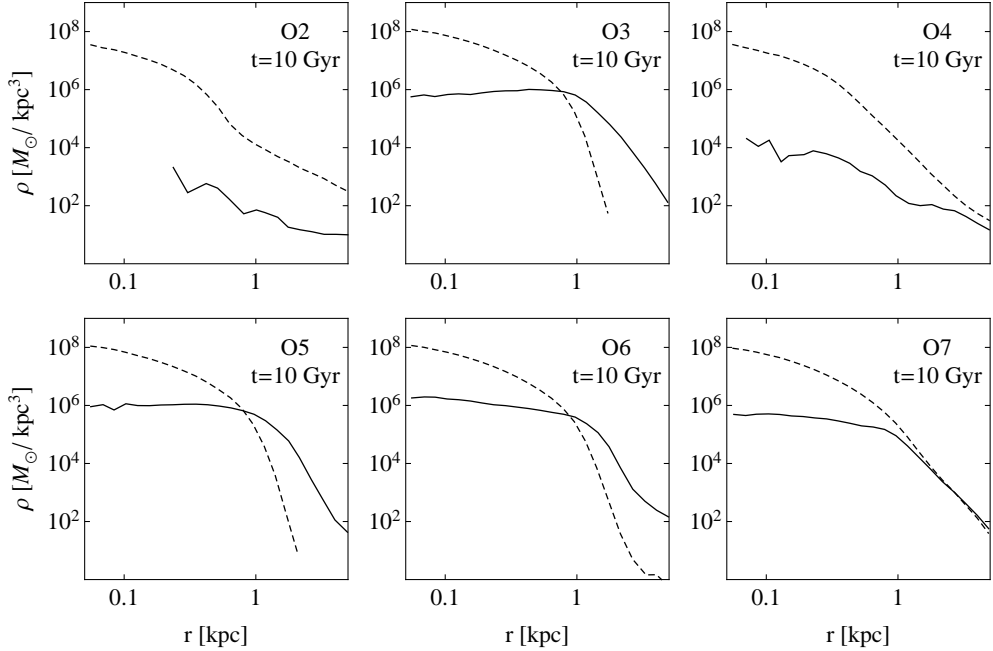}}
\caption{\footnotesize
The density profiles of stars in the inner Population 1 (dashed lines) and the outer Population 2
(solid lines) at the end of evolution (after 10 Gyr) for orbits O2-O7 (from the upper left to the lower right panel).
The populations were initially divided by the half-light radius (as in the upper left panel of Figure~\ref{orbit1}).
}
\label{orbits2-7}
\end{figure*}

\section{Evolution of stellar populations}

In order to see how stellar populations may be affected by tidal evolution we divided the stellar component of the
dwarf galaxy into two populations, each containing initially half of the total number of stars. The inner population
(Population 1) was selected to contain all stars with radii $r < r_{1/2}$ and the outer population (Population 2)
all stars with $r > r_{1/2}$ where $r_{1/2}=0.7$ kpc is the three-dimensional half-light radius of the initial
stellar disk. Although quite simplistic, the division probably reflects to some extent the actual distribution
of stars of different metallicity following from multiple star formation episodes. The stars formed later are likely to
be produced from the gas expelled at earlier stages which was later accreted and sunk to the center of the dwarf galaxy
resulting in the segregation of populations with different stellar properties.

The positions of the stars in each population were traced for the whole evolution time.
Figure~\ref{orbit1} shows the density profiles of the stars in the two populations at the initial time, the subsequent
apocenters and at the final output for orbit O1, which may be considered a typical orbit of dwarf galaxies in the
vicinity of the Milky Way (Diemand et al. 2007; Klimentowski et al. 2010).
Although the stars of the two populations were initially spatially separated,
as the evolution proceeds the outer population tends to populate the inner region forming a core-like distribution, while
the inner population extends to larger radii forming a tail. The stars are lost from both populations due to tidal
forces, as confirmed by the decreasing normalization of the density profiles. For most of the time, Population 1
dominates in the center, and Population 2 in the outer parts. However, with time the outer population
is more heavily stripped and gradually becomes subdominant at all radii. In addition, both components display
a flattening of their respective density profiles at radii $r>1$ kpc signifying the transition to tidal tails.

As expected, the final outcome of the evolution depends sensitively on the orbit on which the dwarf is accreted. We
illustrate this in Figure~\ref{orbits2-7} which shows the final density profiles of the two populations on orbits
O2-O7. The populations were defined initially in exactly the same way as for orbit O1. We see that on the very tight orbit
O2 and the one with the smallest pericenter O4 the outer population is very strongly stripped and very few stars of
this population are still bound to the dwarf. On the other hand, the less tight orbits O3, O5 and O6 allow both
populations to survive and retain their different distributions. The most eccentric orbit O7 is an interesting
intermediate case.

\section{Conclusions}

Using $N$-body simulations of dwarf galaxies orbiting the Milky Way we have studied the evolution of separate
stellar populations in dwarfs. We have shown that in spite of repeated action of tidal forces at pericenters,
the differences in the distribution of stellar populations survive over timescales comparable to the present
age of the Universe. The stars of the inner
stellar component migrate outwards and the ones of the outer one inwards populating the regions initially
devoid of the stars of each type. Still, the populations do not mix completely and
the density profiles of the two components remain distinguishable.
The inner component remains more centrally concentrated and the outer one more extended so that their respective
characteristic radii (e.g. the half-light radii) at present may be very different. In general, the more extended
population is stripped more effectively and thus may become subdominant at all radii for sufficiently tight orbits.

We have demonstrated that the presence of multiple stellar populations in dSph galaxies of the Local Group is
entirely consistent with the tidal stirring scenario which envisions their formation as following from
interactions of late-type progenitors with a Milky Way-like host galaxies. Even with the instability inherent
to the bar formation usually taking place at the first pericenter passage, the separate populations do survive. This
should be the case also for the generic example of such interactions, namely the Sagittarius dwarf, which moves
around the Milky Way on a rather tight orbit and is thus subject to particularly strong tidal effects
resulting in the formation of pronounced tidal tails (Majewski et al. 2003; Law et al. 2005;
{\L}okas et al. 2010). Indeed, the
presence of multiple stellar populations in Sagittarius and its tidal tails is well established observationally
(Siegel et al. 2007; Chou et al. 2007).

\begin{acknowledgements}
This research and EL{\L}'s participation in EWASS 2012 was partially supported by the Polish National Science
Centre under grant NN203580940. KK acknowledges the summer student program at the Copernicus Center
in Warsaw. We are grateful to A. Helmi for the suggestion to investigate the evolution of separate stellar
populations in tidally stirred dwarfs.
\end{acknowledgements}

\bibliographystyle{aa}

\end{document}